\documentclass[conference]{IEEEtran}
\IEEEoverridecommandlockouts
\usepackage{cite}
\usepackage{amsmath,amssymb,amsfonts}
\usepackage{algorithmic}
\usepackage{graphicx}
\usepackage{textcomp}
\usepackage{subfiles}
\usepackage{booktabs}
\usepackage{hyperref}
\usepackage{xcolor}
\usepackage{multirow}
\usepackage[ruled,linesnumbered]{algorithm2e}
\usepackage{array}
\usepackage{amsthm}
\usepackage{float}
\usepackage{graphicx}
\usepackage{subcaption}
\usepackage[compatibility=false]{caption}
\newtheorem{myDef}{Definition}
\newtheorem{myTheo}{Theorem}

\SetKw{Continue}{continue}
\SetKwComment{Comment}{\# }{ }
\def\BibTeX{{\rm B\kern-.05em{\sc i\kern-.025em b}\kern-.08em
    T\kern-.1667em\lower.7ex\hbox{E}\kern-.125emX}}
\begin{document}

\title{Approximate Diverse $k$-nearest Neighbor Search in Vector Database}

\author{\IEEEauthorblockN{Jiachen Zhao}
\IEEEauthorblockA{\textit{The Chinese University of Hong Kong}\\
jczhao23@cse.cuhk.edu.hk}
\and
\IEEEauthorblockN{Xiao Yan}
\IEEEauthorblockA{\textit{Wuhan University}\\
yanxiaosunny@whu.edu.cn}
\and
\IEEEauthorblockN{Eric Lo}
\IEEEauthorblockA{\textit{The Chinese University of Hong Kong}\\
ericlo@cse.cuhk.edu.hk}
}

\maketitle

\begin{abstract}
Approximate $k$-nearest neighbor search (A$k$-NNS) is a core operation in vector databases, underpinning applications such as retrieval-augmented generation (RAG) and image retrieval. In these scenarios, users often prefer diverse result sets to minimize redundancy and enhance information value. However, existing greedy-based diverse methods frequently yield sub-optimal results, failing to adequately approximate the optimal similarity score under certain diversification level. Furthermore, there is a need for flexible algorithms that can adapt to varying user-defined result sizes and diversity requirements.

To address these challenges, we propose a novel approach that seamlessly integrates result diversification into state-of-the-art (SOTA) A$k$-NNS methods. Our approach introduces a progressive search framework, consisting of iterative searching, diversification, and verification phases. Carefully designed diversification and verification steps enable our approach to efficiently approximate the optimal diverse result set according to user-specified diversification levels without additional indexing overhead.

We evaluate our method on three million-scale benchmark datasets, LAION-art, Deep1M, and Txt2img, using latency, similarity, and recall as performance metrics across a range of $k$ values and diversification thresholds. Experimental results demonstrate that our approach consistently retrieves near-optimal diverse results with minimal latency overhead, particularly under medium and high diversity settings.
\end{abstract}

% \begin{IEEEkeywords}
% component, formatting, style, styling, insert
% \end{IEEEkeywords}

\section{Introduction}

Recently, vector databases have emerged as a prominent topic of research and development. A vector database is a specialized system designed to store and manage high-dimensional vectors, often generated by machine learning models, which makes them particularly well-suited for unstructured data, including text, images, and audio~\cite{cao2023knowledgegraphembeddingsurvey, 10.1145/3150226}. Vector databases are widely applied in cross-modality tasks, such as recommendation systems~\cite{Zhao_2024, liu2024multimodalrecommendersystemssurvey}, retrieval-augmented generation (RAG)~\cite{ma2023queryrewritingretrievalaugmentedlarge} and image retrieval~\cite{romero2023zeldavideoanalyticsusing}.

However, the "curse of dimensionality"~\cite{peng2025interpretingcursedimensionalitydistance} renders many traditional $k$-nearest neighbor search ($k$-NNS) methods ineffective in high-dimensional settings. To address this challenge, extensive research has focused on developing efficient index structures that can approximate top-$k$ results, leading to the task known as approximate $k$-nearest neighbor search (A$k$-NNS)~\cite{Huang_2020,xiong2020approximatenearestneighbornegative}. Popular approaches include proximity graphs~\cite{malkov2018efficient,fu2018fast,ssss}, locality-sensitive hashing (LSH)~\cite{andoniOptimalDataDependentHashing2015,weiDETLSHLocalitySensitiveHashing2024}, and product quantization (PQ)~\cite{gao2024rabitqquantizinghighdimensionalvectors,Andr2017}. Among these, proximity graph-based methods have demonstrated outstanding performance and have become the state-of-the-art (SOTA) in empirical evaluations.

% Despite these advances, simply retrieving the top-$k$ results is often insufficient for many modern applications. For example, consider an image retrieval task in which the goal is to find the $k$ most relevant frames from a video. Because consecutive frames are represented by similar embeddings in the vector database, the retrieved set is likely to contain many nearly identical frames, providing limited information to video analysts. This highlights the importance of result diversification: ensuring that the returned results are not only relevant but also diverse, thereby increasing their utility in practical scenarios.

\begin{figure}
  \centering
  \includegraphics[page=2, width=\linewidth]{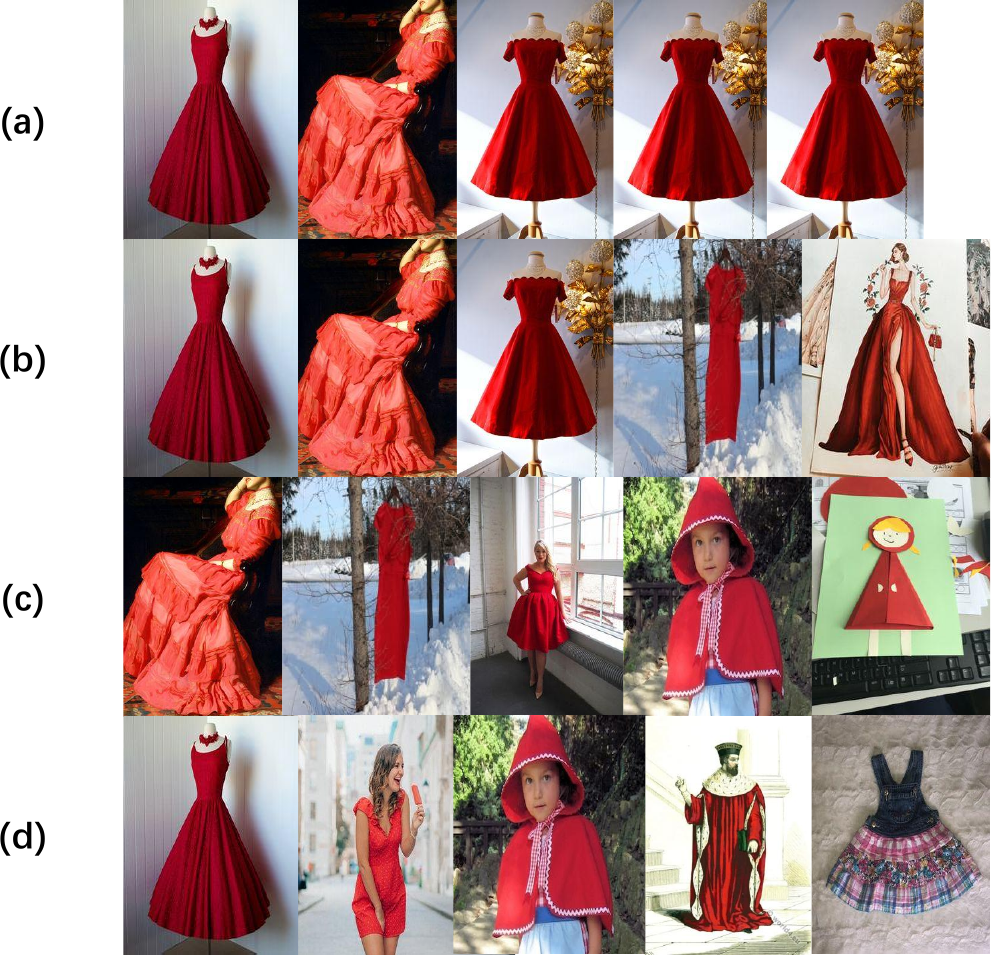}
  \caption{Query results from LAION-art~\cite{schuhmann2022laion5bopenlargescaledataset} dataset. $k=5$ and $q_{text}=$"a photo of a red dress" for all 4 cases. (a) The results without diversification. (b) The optimal results when $\epsilon=0.96$. (c) The optimal results when $\epsilon=0.6$. (d) The results of greedy algorithm when $\epsilon=0.6$.}
  \label{fig:example}
\end{figure}

Despite these advances, simply retrieving the top-$k$ results is often insufficient for many modern applications. For example, consider a recommender system that finds the $k$ most relevant products on an e-commerce website~\cite{9151184} for customers. Products with only slight differences (price, shipping address, or other minor attributes) are represented by similar embeddings in the vector database so the retrieved set is likely to contain many nearly identical results, providing limited information to customers. This highlights the importance of result diversification: ensuring that the returned results are not only relevant but also diverse, thereby increasing their utility in practical scenarios. Figure~\ref{fig:example} illustrates how a diverse set provides more information. The query vector $q$ is the embedding of the sentence "a photo of a red dress", computed using CLIP-ViT-B-32~\cite{radford2021learningtransferablevisualmodels}. We search for the top $k=5$ most similar images in the LAION-art~\cite{schuhmann2022laion5bopenlargescaledataset} dataset. Without result diversification (case (a)), the returned set contains three redundant images, offering no more information than a set with $k=3$. When result diversification is applied (case (b)), these redundant results are effectively excluded from the retrieved set.

Another important consideration is that, for the same dataset and index structure, users may want to adjust diversification level based on different requirements. For example, on e-commerce websites, users may wish to set a low diversification level to exclude only identical products, or a high diversification level to filter out products that differ in attributes such as color or brand. Therefore, in practical applications, it is highly beneficial for diverse search methods to support various user-defined levels of diversification in the result set. The comparison between case (b) and case (c) in Figure~\ref{fig:example} illustrates how the result set varies under different levels of diversification.
% Additionally, professional users—such as video analysts~\cite{romero2023zeldavideoanalyticsusing}—often have in-depth knowledge of the embedding models and the underlying database. By profiling, they can anticipate the results at different levels of diversification.

Inspired by practical needs, the diverse $k$-nearest neighbor search problem is defined as the following:
\begin{myDef}
\label{Def:div}
(\textbf{Diverse $k$-nearest Neighbor Search}) Given a dataset $V = \{v_1, v_2, \ldots, v_N\}$, a similarity function $\mathrm{sim}(u, v)$, and a query $Q = (q, k, \epsilon)$, the \textbf{diverse $k$-nearest neighbor search} (D$k$-NNS) problem aims to find the \textbf{optimal diverse set} $R = \{r_1, r_2, \ldots, r_k\} \subseteq V$ satisfying the following conditions:
\begin{itemize}
    \item \textbf{Diversification condition}: $\forall r_i,r_j\in R$, $i\neq j$, such that $sim(r_i,r_j)<\epsilon$
    \item \textbf{Size condition}: $|R|= k$
    \item \textbf{Optimality condition}: $\sum_{i=1}^{k}sim(r_i,q)$ is maximized
\end{itemize}
\end{myDef}
As long as a vector set satisfies the diversification condition, it is a diverse set that is said to be at a diversification level of $\epsilon$. But only the sets satisfying both the diversification and size conditions are considered valid diverse sets. The similarity function $sim(u, v)$ quantifies how similar two vectors are—the higher its value, the more similar the vectors. Accordingly, we refer to $sim(v, q)$ as the similarity score of $v$, and $\sum_{i=1}^{k} sim(r_i, q)$ as the total similarity score of $R$. Notice that for a dataset of size $N$, the number of valid diverse sets is limited, and therefore the optimal diverse set must exists and is the valid diverse set with the highest total similarity score.

Our problem definition is similar to that in \cite{astar}, except that we allow the user to specify $\epsilon$ in the query $Q$, while \cite{astar} assumes that $\epsilon$ is fixed and known in advance. Lu Qin et al.~\cite{astar} proved that D$k$-NNS is an NP-hard problem and proposed a method that applies the A* algorithm to an index structure called diversity graph to efficiently find the optimal result. Although this method guaranties optimality, it suffers from high search latency on large datasets and is not suitable for modern applications. As a result, in practice, the requirement for optimality is often relaxed to reduce latency, leading to the development of the approximate diverse $k$-nearest neighbor search (AD$k$-NNS) problem.

Many previous works~\cite{chen2018fastgreedymapinference,drosou2013disc} have recognized the importance of diversification and proposed various methods to enhance the diversity of retrieval results. However, most of these approaches rely on greedy algorithms, which iteratively select the vector $v$ with the highest score and remove all vectors close to $v$. While these methods make locally optimal choices at each step, they do not necessarily achieve global optimality, leading to sub-optimal approximations of the true D$k$-NNS objective. The sub-optimality of the greedy algorithm is illustrated in Figure~\ref{fig:example}, case (d). While the locally optimal strategy ensures that the first selection is of high quality, it may leave only poor options in subsequent steps, leading to inferior overall results. For example, the last image in case (d) shows a dress of wrong color, highlighting the loss of overall quality compared to the optimal results in case (c). In addition, to narrow the search scope, many methods~\cite{romero2023zeldavideoanalyticsusing} first identify a subset of predefined size and then diversify the results within this subset. However, for queries that require a high level of diversification, this approach may not be sufficient to produce exactly $k$ results, resulting in a loss of quality. These limitations become especially problematic in scenarios where the quality of the results is critical.

To address these limitations, we propose several techniques to solve the AD$k$-NNS problem. First, to support dynamic queries and handle scenarios where the result set size is initially unknown, we propose progressive search, a modified search algorithm for proximity graphs. Second, to address the sub-optimality of greedy-based approaches and improve result quality, we adopt the div-A* algorithm proposed in \cite{astar}, which guarantees optimal results. To overcome the limitations of div-A* and reduce the search scope, we propose two methods to estimate the minimum candidate queue size required to ensure an optimal diverse set throughout the data set. Finally, we analyze the computational complexity and provide error bounds for the proposed methods.

Our best method consistently achieves near-optimal results across all experimental settings with millisecond-level latency. Under high diversification levels, it delivers both higher result quality and lower latency compared to baseline greedy-based methods. For example for high diversification queries with $k=15$ on the Txt2img dataset, the recall rate improves from 67\% to 92\% while the latency decreases from 190 ms to 81 ms. Notably, our approach does not require any additional index structures, making it highly portable and readily applicable to any A$k$-NNS algorithm.

Our contribution can be summarized as follows:
\begin{itemize}
    \item We identify the limitations of existing methods for the AD$k$-NNS problem, i.e. the sub-optimal result quality and the possibility of returning fewer than $k$ results.
    \item We propose progressive search, a modified beam search algorithm used to handle cases where the result size is unknown at first.
    \item We propose two methods to estimate the candidate number required to derive the optimal diverse set. We also derive the computational complexity and provide error bounds for our methods.
\end{itemize}
\section{Preliminary}\label{related}

\subsection{Proximity Graph Technique}
A$k$-NNS requires the construction of an index during the offline stage to efficiently reduce the number of similarity computations between queries and data vectors during an online search. Among all techniques, proximity graphs have demonstrated superior performance in A$k$-NNS. In a proximity graph, each node representing a vector is connected to $M$ neighboring nodes within close proximity. Among the SOTA proximity graph approaches, many techniques stand out, for example Hierarchical Navigable Small World (HNSW) graphs\cite{malkov2018efficient}, Navigating Spreading-out Graphs (NSG)\cite{fu2018fast}, and Vamana\cite{ssss}. HNSW leverages the concept of small-world networks by constructing a multi-layer graph structure for efficient search. The upper layers of the graph are sparse, enabling rapid navigation towards the vicinity of the query point, while the lower layers are dense to ensure high recall rate. NSG introduces a connectivity test to dynamically adjust the graph structure during construction. Vamana constructs proximity graphs by carefully controlling the number of outgoing edges for each node, ensuring the graph remains sparse without sacrificing search quality. It uses a greedy neighbor selection process that balances proximity and graph connectivity, enabling efficient and scalable approximate nearest neighbor search even for very large datasets.

The search process on a proximity graph typically follows a traversal strategy similar to breadth-first search (BFS), commonly known as beam search. To better introduce beam search and to clearly present our proposed modifications in Section~\ref{methodology}, we provide its pseudo-code in Algorithm~\ref{alg:beam}. At query time, the beam search starts from a designated entry point $s$ (line 1). For each iteration (lines 2-12), the algorithm first explores neighboring nodes connected to the current node, computes its similarity score with respect to the query vector, and maintains them in a candidate queue $C$ (lines 5-7), which is a sorted list with maximum length $L$ (lines 9-11) arranged in decreasing order of similarity score (line 8). The search continues until all nodes in the candidate queue are stable, at which point the algorithm terminates and returns the first $k$ vectors as the approximate nearest neighbors of the query (line 13). Users can adjust the parameter $L$ to balance the trade-off between query latency and accuracy.

\begin{algorithm}[!t]
\caption{BeamSearch$(G,q,k,L)$}
\label{alg:beam}
\KwData{Proximity graph $G$ with start point $s$, query vector $q$, beam width $L$}
\KwResult{$k$ most similar vectors to $q$}
Initialize $C\gets\{s\}$, $p\gets 0$\\
\While{$p<L$}{
$p\gets$ the index of the first unstable candidate in $C$\\
mark $C[p]$ as stable\\
\For{all neighbor $v$ of $C[p]$}{
$C\gets C\cup\{v\}$
}
$C\gets sort(C)$\\
\If{$C.size>L$}{$C\gets C.resize(L)$}
}
\Return $C.resize(k)$ 
\end{algorithm}

In our work, we utilize the HNSW index as the underlying proximity graph, which is widely adopted in libraries such as FAISS~\cite{douze2024faiss}. This choice reflects our focus on advancing the search algorithm itself, rather than proposing improvements to the index structure.

\subsection{Result Diversification}

Numerous studies have recognized the importance of results diversification. In this section, we briefly introduce several representative approaches that have the potential to solve the AD$k$-NNS problem and discuss their limitations.

\subsubsection{A*-based Algorithm}
Lu Qin et al.~\cite{astar} proposed a deterministic method called div-A* that applies the A* algorithm to a pre-built index structure called the diversity graph to solve the D$k$-NNS problem when $\epsilon$ is fixed and known in advance. The diversity graph is defined as follows:
\begin{myDef}\label{dg}
(\textbf{Diversity Graph})
    A diversity graph is an undirected graph $G^\epsilon=(V,E)$ where:
    \begin{itemize}
        \item Each node $v\in V$ is associated with a similarity score $v.score$, computed by $sim(v,q)$.
        \item An edge $(v,v')\in E$ exists if and only if $sim(v,v')>\epsilon$, where $\epsilon$ is the similarity threshold.
    \end{itemize}
\end{myDef}

Notice that node scores must be computed with query $q$, but edge information is only related to the dataset itself. Hence, to apply the div-A* in vector database applications, the scores must be computed during the online stage. According to Definition~\ref{Def:div}, a valid diverse set is an independent set of size $k$ on the diversity graph built with the corresponding $\epsilon$. Thus, the independent set with the highest total score is the optimal diverse set. Finding the optimal solution is an NP-hard problem. However, the div-A* algorithm can significantly reduce the computation using a pruning technique.

During the offline stage, the div-A* algorithm prepares $V$ and $E$ for the diversity graph $G^\epsilon=(V,E)$. During query time, div-A* first computes scores for every node. Then, it does the following: (1) Initialize a tree, where each tree node contains the diverse set of current solution, current total score, and maximum possible total score. (2) Construct the children of the current tree node by adding the next possible vector node to the solution and then computing the current total score and the maximum possible total score. (3) Move to the unexplored child with the highest maximum possible score. Move backward if the size of the solution is already $k$. (4) If the maximum possible score is lower than the largest current score we have seen, then move backward. Otherwise, repeat from Step (2). Figure~\ref{fig:heaptree} is a demonstration of the algorithm on the diversity graph shown in Figure~\ref{sub2}.

The div-A* algorithm is capable of finding the optimal diverse set of size $k$. Moreover, by reusing the search tree, it can efficiently derive the optimal diverse sets for all sizes from 1 to $k-1$ as well.
 
However, this A*-based method has two major limitations: (1) For large $k$, large $N$ and dense $G^\epsilon$, it takes seconds to minutes to handle a single query, which is unacceptable in modern vector database applications. (2) This approach requires a pre-built diversity graph, which only supports a single value of $\epsilon$. To solve the D$k$-NNS problem as defined in Definition~\ref{Def:div}, one must either construct and store multiple diversity graphs for different values of $\epsilon$, or build a new diversity graph at query time. However, both are impractical, since the construction of a diversity graph takes $O(N^2)$ time and storing it requires $O(N^2)$ space. Furthermore, since the user-defined similarity threshold is not discrete, it is infeasible to enumerate all possible values of $\epsilon$.

In Section~\ref{methodology}, we propose our method that safely reduces both construction time and space complexity to $O(K^2)$ and utilizes the A*-based approach for the AD$k$-NNS problem.

\subsubsection{Greedy Algorithm}
A commonly used diversification approach to solve the AD$k$-NNS problem in vector databases is the greedy algorithm~\cite{romero2023zeldavideoanalyticsusing}. First, it retrieves $K$ candidate items using ANNS methods. Then iteratively selects the vector $v$ with the highest similarity score to the query and removes all vectors $v'$ that satisfy $sim(v,v')>\epsilon$, where $\epsilon$ is a diversification threshold.

\begin{figure}[!t]
  \centering
  \begin{minipage}{0.23\textwidth}
		\centering
		\includegraphics[page=1, width=\linewidth]{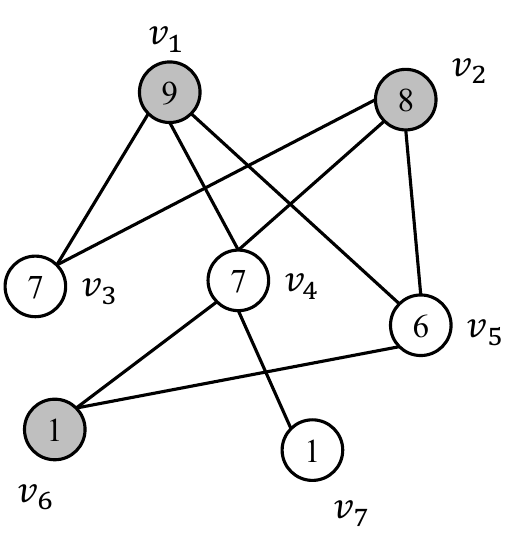}
        \subcaption[]{}
		\label{sub1}
	\end{minipage}
    \begin{minipage}{0.23\textwidth}
		\centering
		\includegraphics[page=2, width=\linewidth]{figures/aabb.pdf}
        \subcaption[]{}
		\label{sub2}
	\end{minipage}
  \caption{An example of a diversity graph illustrating the sub-optimality of the greedy algorithm.
}
  \label{fig:ge}
\end{figure}

\begin{figure}
  \centering
  \includegraphics[page=3, width=\linewidth]{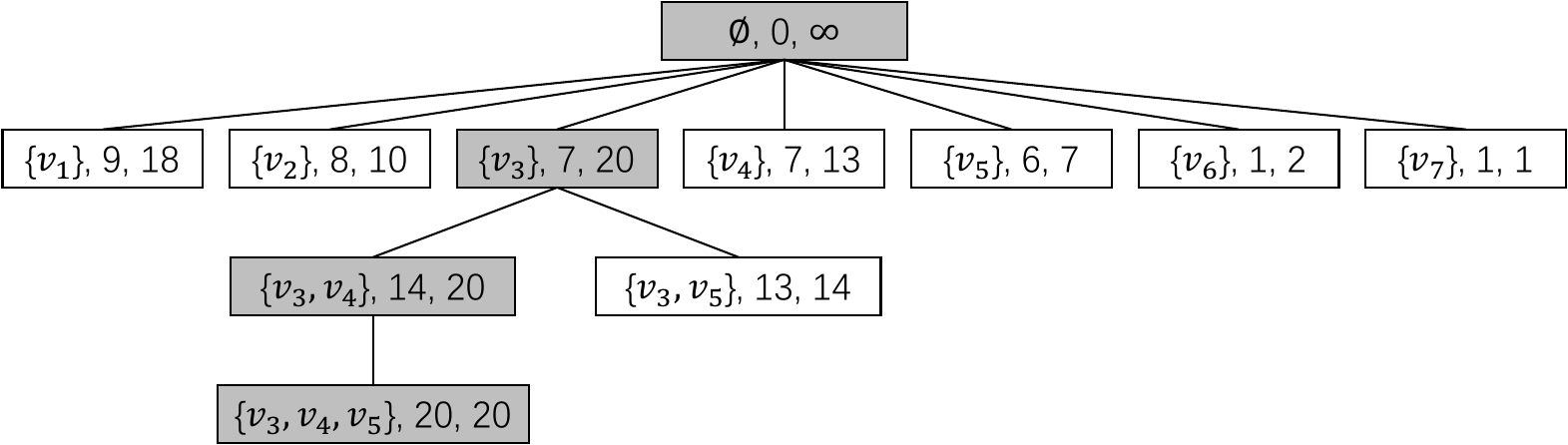}
  \caption{A demonstration of the div-A* algorithm workflow on the diversity graph in Figure~\ref{fig:ge}. The gray nodes indicate the search path taken by the algorithm. The white nodes are pruned from further exploration because their maximum possible scores are lower than the best score observed so far.}
  \label{fig:heaptree}
\end{figure}

Although this method is efficient, it suffers from two key limitations: (1) The use of a fixed candidate set size struggles to accommodate dynamic queries. When $K$ is small, the greedy algorithm can return a diverse set of sizes smaller than $k$, resulting in a significant decrease in quality, measured by the total sum of similarity scores. In contrast, setting $K$ too large introduces unnecessary computational overhead. (2) The greedy algorithm is inherently suboptimal with respect to the total sum of similarity scores in the diversified result set. An example of this sub-optimality is shown in Figure~\ref{fig:ge}. Consider a diversity graph of size $N=7$. When $k=3$, Figure~\ref{sub1} illustrates the result $\{v_1,v_2,v_6\}$ produced by the greedy algorithm, achieving a total score of 18. In contrast, Figure~\ref{sub2} presents the optimal result $\{v_3,v_4,v_5\}$, which attains a total score of 20.

\subsubsection{IP-Greedy}
Kohei Hirata et al.~\cite{10.1145/3523227.3546779} propose a method, further improved by \cite{huang2024diversityaware}, that introduces an adjustable diversification parameter into the similarity function for greedy selection, addressing the diversity-aware $k$-maximum inner product search ($k$-MIPS) problem. The similarity score of a set is defined as:
\begin{equation}
    f(S)=\frac{\lambda}{k}\sum_{\textbf{p}\in S} \textbf{p}\cdot \textbf{q}+c(1-\lambda)\min_{\textbf{p},\textbf{p'}\in S}dist(\textbf{p},\textbf{p'})
\end{equation}
where $c$ is a scaling parameter that adjusts the scale of the distance function to match that of the inner product space. Consequently, the similarity score of a single vector during the query time is defined as follows:
\begin{equation}
    f(\textbf{p},S)=\lambda \textbf{p}\cdot \textbf{q}+c(1-\lambda)\min_{\textbf{p}_a,\textbf{p}_b\in S\bigcup\{\textbf{p}\}}dist(\textbf{p}_a,\textbf{p}_b)
\end{equation}
At each step, the vector with the highest similarity score is chosen, which balances similarity to the query and diversity of previously selected vectors. The process terminates when $k$ diverse vectors have been obtained. To further enhance efficiency, this work employs early-stop techniques to reduce search space.

By redefining the distance function, the IP-greedy algorithm can be adapted for use in vector databases that utilize inner product or cosine similarity as their metric space. However, the approach of Kohei Hirata et al. does not establish a clear relationship between the diversification parameter $\lambda$ and the actual diversification level $\epsilon$ of the set of results. Moreover, through our experiments, we find that the change of $\lambda$ does not necessarily result in significant change the diversity of the output set, measured by $\max_{\textbf{p},\textbf{p'}\in S}sim(\textbf{p},\textbf{p'})$. If any, the decrease in $\lambda$ even make the total similarity smaller. As a result, this design is less suitable for the AD$k$-NNS problem where the user-defined diversification level is essential. This limitation motivates our preference for directly controlling the diversification level rather than relying on an indirect parameter.

\subsection{Design Goal}
We summarize the limitations of existing methods as follows:

 \textbf{User-defined diversification level}: (1) Some approaches lack the ability to directly control the diversification level of the result set. (2) Some approaches only support a fixed set of predefined diversification levels.
 
 \textbf{Sub-optimality}: (1) Greedy-based methods are inherently unable to produce optimal results for all queries. (2) When the pre-defined candidate size is not large enough, the algorithm may return result sets smaller than the desired size, leading to an even greater loss of quality in terms of the total sum of similarity scores.
 
 \textbf{High latency}: (1) The algorithm capable of producing optimal results incurs high runtime costs. (2) When the predefined size of the candidate set is large, the algorithm wastes computational resources on easy queries.

To address these limitations, our aim is to develop an algorithm that supports the following features.
\begin{itemize}
    \item Accepts various user-defined similarity thresholds $\epsilon$.
    \item Returns a diverse set of exactly the size $k$.
    \item Returns a diverse set that is optimal with high probability.
    \item Achieves low latency.
\end{itemize}

\section{Progressive Search} \label{methodology}

In this section, all proposed methods are based on the proximity graph structure introduced in Section~\ref{related}, and we assume that the proximity graph is the only index structure built during the offline stage. Additionally, we make an important assumption that the first $K$ stable candidates in the candidate queue are always accurate, meaning the recall rate of the beam search is 100\%. Although this assumption does not always hold in real-world applications, it is possible to achieve a recall rate approaching 100\% by using a better proximity graph or increasing the size of the candidate queue. This assumption is also supported by \cite{288618}, which shows that the similarity scores of subsequently added candidates decrease almost monotonically after a certain point.

As discussed in Section~\ref{related}, the widely adopted greedy approach treats the beam search process as a black box, applying diversification only to the $K$ results obtained from a predefined beam width. However, this method is inadequate for handling dynamic queries. To address this limitation, we propose a modified beam search framework, called progressive search. The main idea is to make $K$ a variable that is adjusted during the search process, rather than fixing it offline. Initially, $K$ is set to $k$, since at least $k$ candidates are required for a valid diverse set. Once $K$ stable candidates have been identified, we pause the search to check whether these candidates can produce a valid diverse set. If not, we increase $K$ and resume the search. In this way, the progressive search can quickly resolve easy queries while also sufficiently addressing more difficult ones.

To implement the progressive search framework, we introduce two major modifications to the beam search:
\begin{enumerate}
    \item \textbf{Dynamic Beam Width}: Since the final size $K$ of the beam search result is unknown in advance, we eliminate the conventional beam width limit $L$ in Algorithm~\ref{alg:beam} lines 9-11. Instead, we modify Algorithm~\ref{alg:beam} line 2 to make the search terminate when the first $K \times ef$ candidates become stable, where $ef$ is a parameter controlling the efficiency level. This design is consistent with the HNSW\cite{malkov2018efficient} approach to approximate nearest neighbor search (ANNS), where the convergence of a single candidate is determined by the stability of the $ef$ candidate neighbors.
    
\item\textbf{Candidate Queue Reuse}: Because $K$ increases incrementally and the beam search may be invoked multiple times, instead of initiating a new candidate queue in Algorithm~\ref{alg:beam} line 1, we minimize computational overhead by maintaining the candidate queue across search iterations. Each time, the beam search resumes using the candidate queue from the previous iteration.
\end{enumerate}
This modified beam search is named \textbf{progressive beam search} and is used in algorithm descriptions as the function \texttt{ProgressiveBeamSearch}. Note that progressive beam search, as a nearest neighbor search strategy, may perform worse than beam search for the A$k$-NNS problem. However, it is well-suited for the AD$k$-NNS problem to produce diverse sets of higher quality. In the following subsections, we introduce diversification methods designed to work within the progressive search framework. In the last subsection, we analyze the complexity of three proposed algorithms.

\subsection{Greedy-based Search}\label{prosearchsec}
\begin{algorithm}[!t]
\caption{ProgressiveGreedySearch($G,Q,ef$)}\label{alg:progressive}
\KwData{Proximity graph $G$ with start point $s$, query $Q=(q,k,\epsilon)$, efficiency level $ef$}
\KwResult{Diverse set $R$}
Initialize $C\gets\{s\}$, $R\gets\emptyset$, $K\gets k$

\While{R.size $<$ k}{
  $C\gets$ProgressiveBeamSearch$(C,G,q,K,ef)$\\
  \For{candidate $u$ of the first $K$ candidates in $C$}{
  $div\gets True$\\
  \For{vector v in R}{
      \If{$\text{sim}(v,u)>\epsilon$}{
      $div\gets False$
      }
  }
  \If{div=True}{
  $R\gets R\cup\{u\}$
  }
  }
  $K\gets K+k$
}
\Return $R$
\end{algorithm}

The greedy algorithm introduced in Section~\ref{related} can be directly integrated into the progressive search framework. We refer to this approach as \textbf{progressive greedy search} (PGS), detailed in Algorithm~\ref{alg:progressive}. The algorithm initializes both the candidate queue and the result set $R$ as empty sets (line 1). Then it invokes progressive beam search to obtain $K$ stable candidates (line 3). During the pause phase (lines 4–14), the greedy algorithm is applied to the first $K$ stable candidates, adding the qualified ones to $R$ (lines 11–13). Since the greedy algorithm ensures that $R$ is a diverse set, only the size condition needs to be checked (line 2) to make $R$ a valid diverse set. If this condition is not met, the PGS increases $K$ by $k$ (line 15) and begins the next iteration.

Although this method is capable of producing exactly $k$ results and increasing the efficiency level $ef$ can further improve the result quality, it has a fundamental sub-optimality. Therefore, in subsequent methods, we further explore the use of the div-A* algorithm.

\subsection{Div-A*-based Search}
\begin{figure}
  \centering
  \includegraphics[page=8, width=0.8\linewidth]{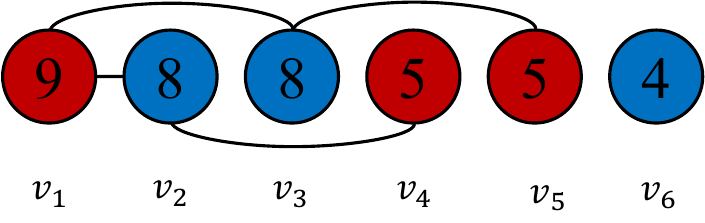}
  \caption{An example illustrating the difference between greedy algorithm and div-A* algorithm.}
  \label{fig:increase}
\end{figure}

The div-A* algorithm, as a result diversification method, can also be integrated into the progressive search framework. However, unlike the greedy algorithm, div-A* can yield better results with a larger candidate queue. For example, suppose $k=3$ and the candidate queue forms the diversity graph shown in Figure~\ref{fig:increase}, both greedy and div-A* algorithms would return a diverse set $\{v_1,v_4,v_5\}$ when $K=5$. However, the optimal diverse set includes $v_6$, the vector with the sixth highest similarity score. In this case, div-A* can return a higher-quality diverse set $\{v_2,v_3,v_6\}$ when the candidate queue size is increased to $K=6$. In contrast, the greedy algorithm would return the same diverse set as with $K=5$, regardless of any further increase in the candidate queue size beyond 5. This difference makes it important to establish an additional stopping condition for div-A*-based methods. We propose two different approaches to estimate the final value of $K$, along with the corresponding theorems to ensure optimality in the following subsections.

\subsubsection{Degree-based Estimation}
\begin{algorithm}[!t]
\caption{ProgressiveDegreeSearch($G,Q,ef$)}\label{alg:degree}
\KwData{Proximity graph $G$ with start point $s$, query $Q=(q,k,\epsilon)$, efficiency level $ef$}
\KwResult{Diverse set $R$}
Initialize $C\gets\{s\}$, $K\gets k$

\While{the first $K$ candidates are not all stable}{ 
  $C\gets$ProgressiveBeamSearch$(C,G,q,K,ef)$\\ 
  $G^\epsilon\gets$ BuildDiversificationGraph$(C.resize(K),\epsilon)$\\
  $\Phi\gets k-1$ nodes with highest degree in $G^\epsilon$\\
  $K\gets\sum_{v\in\Phi}(\phi_v+1)+1$ 
}
$R\gets$div-A*$(G^\epsilon, k)$\\
\Return $R$
\end{algorithm}
\begin{figure}[!t]
  \centering
  \begin{minipage}{0.45\textwidth}
		\centering
		\includegraphics[page=5, width=0.8\linewidth]{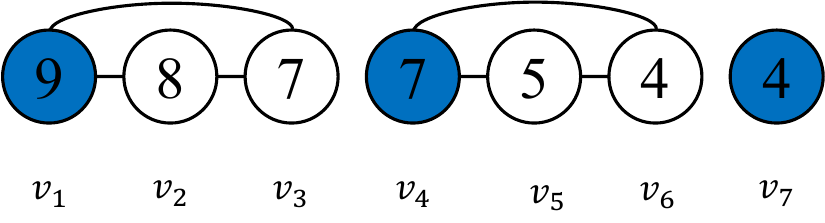}
        \subcaption[]{}
		\label{Proof_1}
	\end{minipage} \\
    \begin{minipage}{0.45\textwidth}
		\centering
		\includegraphics[page=6, width=0.9\linewidth]{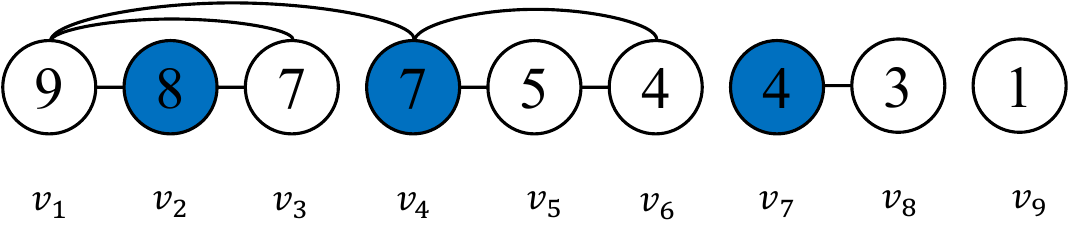}
        \subcaption[]{}
		\label{Proof_2}
	\end{minipage}
  \caption{Examples illustrating Theorem~\ref{Theo:degree} for $k=3$ on a diversity graph.}
  \label{fig:degreeexp}
\end{figure}

Motivated by the concept of the diversity graph, we propose a method that utilizes the degree of nodes within the diversity graph to estimate the number $K$ of candidates required to compute the optimal result set. Suppose $V_K\subset V$ is a set that satisfies $|V_K|=K$ and $\forall v\in V_K$, $\forall u\in V\setminus V_K$, we have $sim(v,q)\geq sim(u,q)$. $G^\epsilon=(V_K,E)$ is a diversity graph built with $V_K$. Let $\Phi\subset V_K$, $|\Phi|=k-1$ be the set of nodes with the highest degree in $G^\epsilon$, i.e., $\forall v\in\Phi$, $\forall u\in V_K\setminus\Phi$, we have $\phi_v\geq \phi_u$, where $\phi_v$ is the degree of $v$ in $G^\epsilon$. Based on this, we can establish the following theorem:

\begin{myTheo}
\label{Theo:degree}
If $K\geq\sum_{v\in\Phi}(\phi_v+1)+1$, then $V_K$ is sufficient to find the optimal diverse set of $V$.
\end{myTheo}

\begin{proof}
The diversification condition in Definition~\ref{Def:div} indicates that $R$ is an independent set of $G^\epsilon$.

To satisfy the size condition in Definition~\ref{Def:div}, we demonstrate that $K$ vectors are sufficient to produce a diverse set of size $k$ even in the worst case scenario, where every group of $\phi_v + 1$ nodes in the diversity graph forms a fully connected subgraph (clique). In this setting, selecting one node from such a subgraph precludes the selection of any other nodes from the same subgraph due to the mutual connections. Therefore, to construct a diverse set of size $k$, we require $k-1$ disjoint cliques, each of size $\phi_v + 1$, along with one additional node. Consequently, the total number of nodes required is $\sum_{v\in\Phi}(\phi_v+1)+1$.

To show that the optimality condition is satisfied, we prove it by contradiction. Assume that the optimal diverse set contains $k_{out}\geq 1$ vectors that are from $V\setminus V_K$. Suppose that the rest $k-k_{out}$ vectors form a set $\Phi'\subset\Phi$ such that $|\Phi'|=k-k_{out}$ and $\forall v\in\Phi'$, $\forall u\in \Phi$, we have $\phi_v\geq \phi_u$. Then they are connected to at most $\sum_{v\in\Phi'}\phi_v$ vectors. Even if all connected vectors are in $V_K$, there are $K-\sum_{v\in\Phi'}(\phi_v+1)=1+\sum_{v\in\Phi\setminus\Phi'}(\phi_v+1)\geq 1$ vectors that can be selected in the independent set and are closer to the query vector $q$. Hence, swapping one of them with one of the $k_{out}$ vectors leads to a better solution. By contradiction, the nearest $K=\sum_{v\in\Phi}(\phi_v+1)+1$ vectors is sufficient to obtain $R$.
\end{proof}

A demonstration of Theorem~\ref{Theo:degree} is shown in Figure~\ref{fig:degreeexp}.

Leveraging Theorem~\ref{Theo:degree}, we design an algorithm called \textbf{progressive degree search} (PDS), as detailed in Algorithm~\ref{alg:degree}. Before applying the div-A* algorithm to improve result quality, PDS attempts to obtain a sufficient number of candidates that satisfy the conditions specified in Theorem~\ref{Theo:degree}. In each iteration (lines 3–6), PDS invokes the progressive beam search to produce a candidate queue of size $K\times ef$. Subsequently, it constructs a diversity graph $G^\epsilon$ from the first $K$ candidates and identifies the $k-1$ nodes with the highest degree. To enhance efficiency, PDS incrementally updates the diversity graph from the previous iteration, modifying only the newly discovered nodes from the latest progressive beam search. Next, PDS updates $K$ according to Theorem~\ref{Theo:degree}, using $K = \sum_{v \in \Phi} (\phi_v + 1) + 1$. The loop terminates only when the number of stable candidates is greater than or equal to the current $K\times ef$. Finally, PDS applies the div-A* algorithm to the candidate set and returns the results (lines 8–9).

Since the degree information required for this approach can be precomputed during an offline phase, PDS is able to quickly construct the diversity graph and compute the candidate size $K$. In addition, the computationally expensive div-A* algorithm needs to be run only once.

However, PDS suffers from two limitations: (1) Computing and storing accurate degree information offline incurs a time complexity of $o(N^2)$ and requires extra index storage, so we do not apply such optimization in the experiments. (2) In high diversification settings, since Theorem~\ref{Theo:degree} provides only a sufficient condition, the resulting value of $K$ is often unnecessarily large. An example is shown in Figure~\ref{fig:degreeexp}. Two figures share the same query vector but differ in the value of $\epsilon$. In each case, the blue nodes represent the optimal diverse set. Figure~\ref{Proof_1} demonstrates the scenario in which every subgraph forms a fully connected clique. In this case, Theorem~\ref{Theo:degree} provides the tightest estimate $K=7$, since the optimal result must include node $v_7$. However, if an extra edge, specifically $(v_1, v_4)$, is added to the diversity graph in Figure~\ref{Proof_2}, Theorem~\ref{Theo:degree} then gives an overestimate of $K=9$, even though only seven nodes are needed to compute the optimal result.

\subsubsection{Score-based Estimation}

\begin{algorithm}[!t]
\caption{ProgressiveScoreSearch($G,Q,ef$)}\label{alg:score}
\KwData{Proximity graph $G$ with start point $s$, query $Q=(q,k,\epsilon)$, efficiency level $ef$}
\KwResult{Diverse set $R$}
Initialize $C\gets\{s\}$, $K\gets k$\\

$C,K\gets$ProgressiveGreedySearch*($G,Q,ef$)

$minValue\gets -\infty$

\While{$minValue< C[K].score$}{
\If{not first loop}{
$K,C\gets$ProgressiveBeamSearch*$(C,G,q,minValue,ef)$}
$G^\epsilon\gets$ BuildDiversificationGraph$(C.resize(K),\epsilon)$\\
$R_1,R_2,...,R_k\gets $div-A*$(G^\epsilon, k)$\\
$minValue\gets \min_{0<i<k}\{\frac{R_k.score-R_i.score}{k-i}\}$
}
\Return $R_k$
\end{algorithm}

To address the limitations of PDS, we propose a score-based method in this subsection. This approach leverages score information, which can be obtained during progressive beam search, to provide a tighter approximation of $K$ and improve overall efficiency.

We designate the first $K$ elements in the candidate queue $C$ as $C_K=\{r_1, r_2, \ldots, r_{K}\}$ arranged in descending order of similarity scores $s_1,s_2,...,s_K$, which are computed by $s_i=sim(r_i,q)$, and we assume that these elements precisely constitute the solution to the $K$ nearest neighbors of $q$. Assume $C_K$ is capable of deriving a diverse set of size $k$ from the div-A* algorithm, then we can acquire the optimal result sets ranging in size from 1 to $k$: $R_1,R_2,...,R_k$ and their total scores $S_1,S_2,...,S_{k}$. Consequently, the following theorem can be formulated.

\begin{myTheo}
\label{Theo:score}
    If $\min_{0<i<k}\{\frac{S_k-S_i}{k-i}\}>s_K$, then the current solution is optimal with total score $S_k$.
\end{myTheo}

\begin{figure}
  \centering
  \includegraphics[page=4, width=0.8\linewidth]{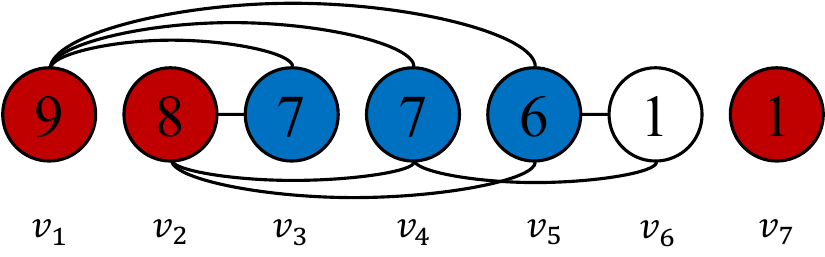}
  \caption{An example of Theorem~\ref{Theo:score} for $k=3$ on a diversity graph.}
  \label{fig:score}
\end{figure}

\begin{proof}
We prove by contradiction. If $\min_{0<i<k}\{\frac{S_k-S_i}{k-i}\}>s_K$ is satisfied but $R_k$ is not the optimal solution, then the actual optimal result set $R'_k$ must contain vectors that are not in the first $K$ candidates. Suppose that there are $j$ vectors that have index larger than $K$, then the total score of $R'_k$ satisfies the following inequalities:
\begin{equation}
    S_k<S'_k\leq S_{k-j}+s_K*j
\end{equation}
However, it contradicts the assumption that $\frac{S_k-S_{k-j}}{j}>s_K$. Hence, the assumption is wrong and Theorem~\ref{Theo:score} is proved.
\end{proof}
A demonstration of Theorem~\ref{Theo:score} is shown in Figure~\ref{fig:score}. Among the first $K=6$ nodes, the blue nodes constitute the current optimal result set $R_3 = \{v_3, v_4, v_5\}$, while the two red nodes form the optimal result set of size 2, $R_2 = \{v_1, v_2\}$. The corresponding total scores are $S_2 = 17$ and $S_3 = 20$, so condition $s_6 < S_3 - S_2$ is satisfied. In this case, the best possible diverse set of size 3 that includes nodes with index greater than $K$ is $\{v_1, v_2, v_7\}$, but this set achieves a smaller score than $S_3$. Therefore, $R_3$ is guaranteed to be the optimal diverse set over the entire dataset.

Theorem~\ref{alg:score} can be applied as a strong early stopping criterion. However, to utilize it effectively, it is essential that $C_K$ can derive a diverse set of size $k$. To address cases where this assumption may not hold, we integrate PGS within our algorithm.

The \textbf{progressive score search} (PSS) algorithm is presented in Algorithm~\ref{alg:score}. In the first phase (line 2), PSS utilizes the PGS routine described in Algorithm~\ref{alg:progressive}. Instead of immediately returning the $k$ results, PSS preserves the current candidate queue $C$ to further improve result quality. For each iteration (lines 5–10), PSS first obtains $K\times ef$ candidates. The modified progressive beam search procedure, \texttt{ProgressiveBeamSearch*} (line 6), terminates when the score of the last stable candidate drops below \texttt{minValue} and returns the current number $K\times ef$ of stable candidates. PSS then constructs the diversity graph using the first $K$ candidates (line 8), incrementally updating the graph from the previous iteration as in PDS. Next, PSS applies the div-A* algorithm (line 9). Since the greedy phase ensures the existence of a result set of size $k$, div-A* can derive result sets of size ranging from 1 to $k$. The value $minValue$ is then computed as $minValue = \min_{0 < i < k} \{ \frac{S_k - S_i}{k-i} \}$, which serves as an early stopping criterion. If the condition $minValue > C[K].\texttt{score}$ holds, the algorithm terminates and returns the current $R_k$ as the result set. Otherwise, it proceeds to the next iteration until the last stable candidate's score falls below $minValue$.

\begin{figure}
  \centering
  \includegraphics[page=7, width=0.9\linewidth]{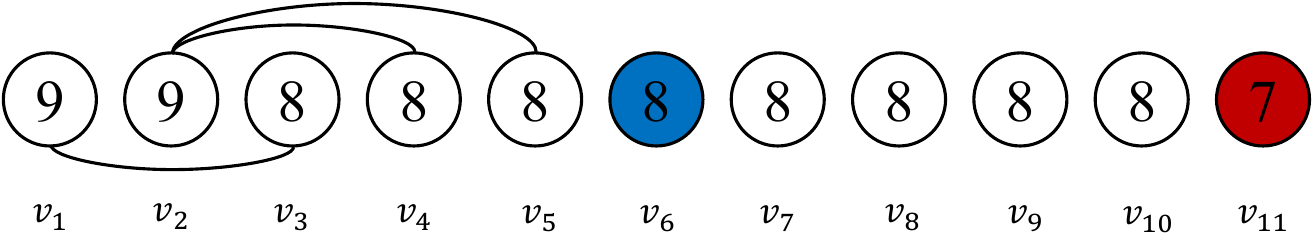}
  \caption{A case in which Theorem~\ref{Theo:score} suffers from overestimation. $K_{PSS}=11$ (the red node) and $K_{PDS}=6$ (the blue node).}
  \label{fig:comp}
\end{figure}

Similarly to degree-based estimation, Theorem~\ref{Theo:score} also provides a sufficient condition and may give an overestimation in certain cases. An example is shown in Figure~\ref{fig:comp}. The optimal diverse sets of size $k=2,3$ computed from the first 10 nodes are $R_2=\{v_1,v_2\}$ with $S_2=18$ and $R_3=\{v_1,v_4,v_5\}$ with $S_3=25$. To satisfy the condition in Theorem~\ref{Theo:score}, the estimation given by PSS would be $K_{PSS}=11$. However, only the first 6 nodes are necessary to compute the optimal diverse set. It is worth mentioning that PDS gives the best estimate of $K_{PDS}=6$.The estimation provided by PDS is much tighter for two main reasons: (1) PDS performs well when the diversity graph is sparse, and (2) PSS tends to perform poorly when multiple nodes have the same similarity score. In practice, since the vectors in real-world datasets are embeddings computed by deep learning models, they are typically diverse to maximize activation. As a result, the similarity scores of nodes are generally distinct, making PSS an efficient method in real applications.

\subsection{Analysis}\label{Analysis} 
\subsubsection{Complexity}
\textbf{Indexing:} All three algorithms employ HNSW as the index structure, resulting in identical construction complexity of $O(efMN\log{N})$.

Our approach focuses exclusively on the algorithmic aspect, rather than the indexing component. Thus, only the original HNSW structure is required to support our methods. Consequently, the overall space complexity is $O(MN\log N)$.

\textbf{Runtime: }We utilize Theorem~\ref{Theo:degree} to approximate $K$ in our theoretical analysis, as it provides a sufficient condition. The average degree of nodes in the diversity graph $G^\epsilon(V,E)$ can be expressed as a function of the diversification threshold, denoted $\phi(\epsilon)$. Consequently, $K = O(\phi(\epsilon)k)$.

The search complexity for all approaches is $O(efMK\log{N})$. The greedy diversification component requires $O(Kk)$ time. For the algorithm based on div-A*, constructing the diversity graph incurs a complexity of $O(K^2)$. The div-A* diversification process has a worst-case time complexity of $O(K^k)$. However, this is an overestimate since the algorithm employs early stopping in practice.

\subsubsection{Error bound}
In this section, we discuss the relationship between the recall rate of the A$k$-NNS problem and the DA$k$-NNS problem. Suppose that, under a certain $ef$ setting, the recall rate of the A$k$-NNS problem is $Recall=1-\lambda$. Then the following theorem can be derived:

\begin{myTheo}
    The recall rate of the PDS or PSS, noted as $Recall_{P}$, has a lower bound of $Recall_{P}\geq(1-\frac{K\lambda}{K-k+1})^k$.
\end{myTheo}

\begin{proof}
 In a low diversification setting where $\phi(\epsilon)\approx1$ and the diversity graph is sparse, the absence of a node in the candidate queue does not significantly affect the results. Hence, the recall rate of DA$k$-NNS would be $Recall_{P}=1-\lambda$.

    In highly diversified settings, where the diversity graph is both large and dense, the absence of a single node from the candidate queue can significantly affect the quality of the results. However, if all nodes belonging to the optimal diverse set are present in the retrieved candidates, the recall rate achieves its maximum value of 1. Let $P$ denote the probability that the optimal diverse set is completely contained within the first $K$ retrieved candidates. This probability $P$ can be formally expressed as follows:
    \begin{equation}
        P=\frac{\tbinom{K-k}{K\lambda}}{\tbinom{K}{K\lambda}}=\frac{\prod_{i=0}^{k-1}(K-K\lambda-i)}{\prod_{i=0}^{k-1}(K-i)}\geq(1-\frac{K\lambda}{K-k+1})^k
    \end{equation}
    Therefore, the lower bound of $Recall_{P}$ is given by $(1-\frac{K\lambda}{K-k+1})^k$, since $P$ represents the probability that the recall rate is 1.
    
\end{proof}

\subsubsection{Limitation}
For the same proximity graph and entry point, the search paths of both the beam search and the progressive beam search should be identical. However, the progressive beam search typically requires more time to retrieve the same number of stable candidates compared to the beam search, as demonstrated in the performance evaluation section. This increased latency is due to the higher cost of maintaining the candidate queue in progressive beam search. The candidate queue functions as a priority queue, and without a fixed beam width, its size can grow significantly, making insert operations expensive. Moreover, since we lack sufficient information to discard candidates during the search process, it is challenging to limit the queue size without compromising the quality of the results. This limitation becomes particularly problematic when applying the progressive search framework to larger datasets—such as billion-scale datasets—which generally require longer search paths and result in even larger candidate queues.
\section{Performance Evaluation}
\subsection{Experiment Setting}

\begin{table}[]
\caption{Dataset information}
\label{tab:dataset}
\begin{tabular}{lllll}
\toprule
Dataset & Dimension & Size & Space  &Query size   \\
\midrule
Deep1M~\cite{Babenko_2016_CVPR}  & 256       & 1M   & Euclidian     &1,000\\
Txt2img~\cite{pmlr-v176-simhadri22a} & 200       & 1M   & Inner product &1,000\\
LAION-art~\cite{schuhmann2022laion5bopenlargescaledataset}   & 512       & 1M   & Cosine     &300  \\
\bottomrule
\end{tabular}
\end{table}

\begin{table*}[]
\centering
\begin{tabular}{@{}lllllll@{}}
\toprule
\textbf{Dataset} & \textbf{Method} & $k=10,\phi(\epsilon)< 10$ & $k=10,\phi(\epsilon)\approx 100$ & $k=10,\phi(\epsilon)\approx 500$ & $k=5,\phi(\epsilon)\approx 500$ & $k=15,\phi(\epsilon)\approx 500$ \\ \midrule
\multirow{5}{*}{Deep1M}    & Div-A*         &   \textbf{L:} 820, \textbf{S:} 3.5139   &   \textbf{L:} 835, \textbf{S:} 3.5013    &   \textbf{L:} 968, \textbf{S:} 3.3974    &   \textbf{L:} 933, \textbf{S:} 1.7813    &   \textbf{L:} 2399, \textbf{S:} 4.9465   \\
                           & Greedy\_400 &   \textbf{L:}2,\textbf{S:}3.5118,\textbf{R:}98.7    &   \textbf{L:}2,\textbf{S:}3.4974,\textbf{R:}97.3    &   \textbf{L:}2,\textbf{S:}3.3680,\textbf{R:}83.2    &   \textbf{L:}2,\textbf{S:}1.7708,\textbf{R:}84.1    &   \textbf{L:}2,\textbf{S:}4.8751,\textbf{R:}82.1    \\
                           & PGS            &   \textbf{L:}12,\textbf{S:}3.5128,\textbf{R:}99.1   &   \textbf{L:}15,\textbf{S:}3.4985,\textbf{R:}97.8    &  \textbf{L:}56,\textbf{S:}3.3741,\textbf{R:}83.9     &   \textbf{L:}17,\textbf{S:}1.7699,\textbf{R:}83.7    &   \textbf{L:}57,\textbf{S:}4.9125,\textbf{R:}82.9    \\
                           & PDS            &   \textbf{L:}36,\textbf{S:}3.5129,\textbf{R:}99.1   &   \textbf{L:}151,\textbf{S:}3.5007,\textbf{R:}98.6    &   N/A   &   \textbf{L:}1654,\textbf{S:}1.7818,\textbf{R:}77.1    &   N/A    \\
                           & PSS            &   \textbf{L:}37,\textbf{S:}3.5133,\textbf{R:}99.1   &   \textbf{L:}41,\textbf{S:}3.5003,\textbf{R:}99.1    &   \textbf{L:}66,\textbf{S:}3.3951,\textbf{R:}98.0    &   \textbf{L:}27,\textbf{S:}1.7782,\textbf{R:}96.1    &    \textbf{L:}137,\textbf{S:}4.9424,\textbf{R:}98.2   \\\midrule
\multirow{5}{*}{LAION-art} & Div-A*         &   \textbf{L:} 800, \textbf{S:} 3.1669   &   \textbf{L:} 811, \textbf{S:} 3.1487    &   \textbf{L:} 1812, \textbf{S:} 3.1254    &   \textbf{L:} 1494, \textbf{S:} 1.5994    &   \textbf{L:} 5238, \textbf{S:} 4.6186    \\
                           & Greedy\_400 &   \textbf{L:}2,\textbf{S:}3.1567,\textbf{R:}84.7    &   \textbf{L:}2,\textbf{S:}3.1343,\textbf{R:}75.1    &   \textbf{L:}2,\textbf{S:}3.0950,\textbf{R:}66.9    &   \textbf{L:}2,\textbf{S:}1.5902,\textbf{R:}72.4    &   \textbf{L:}2,\textbf{S:}4.5129,\textbf{R:}64.5   \\
                           % & IP-greedy      &       &       &       &       &       \\
                           & PGS            &   \textbf{L:}17,\textbf{S:}3.1594,\textbf{R:}87.5    &    \textbf{L:}13,\textbf{S:}3.1365,\textbf{R:}77.6   &    \textbf{L:}103,\textbf{S:}3.1100,\textbf{R:}73.8   &    \textbf{L:}17,\textbf{S:}1.5913,\textbf{R:}73.2   &    \textbf{L:}160,\textbf{S:}4.5955,\textbf{R:}74.4  \\
                           & PDS            &   \textbf{L:}5861,\textbf{S:}3.5007,\textbf{R:}81.9    &  N/A     &   N/A    &   N/A    &   N/A    \\
                           & PSS            &   \textbf{L:}47,\textbf{S:}3.1596,\textbf{R:}88.3    &    \textbf{L:}39,\textbf{S:}3.1390,\textbf{R:}86.3   &   \textbf{L:}97,\textbf{S:}3.1144,\textbf{R:}85.0    &    \textbf{L:}45,\textbf{S:}1.5935,\textbf{R:}83.4   &   \textbf{L:}366,\textbf{S:}4.6046,\textbf{R:}85.3    \\\midrule
\multirow{5}{*}{Txt2img}   & Div-A*        &   \textbf{L:} 371, \textbf{S:} 4.8280   &   \textbf{L:} 383, \textbf{S:} 4.7775    &   \textbf{L:} 435, \textbf{S:} 4.6755    &   \textbf{L:} 378, \textbf{S:} 2.4028    &   \textbf{L:} 1157, \textbf{S:} 6.8896    \\
                           & Greedy\_400 &   \textbf{L:}1,\textbf{S:}4.8255,\textbf{R:}96.1    &   \textbf{L:}1,\textbf{S:}4.7653,\textbf{R:}84.5    &   \textbf{L:}1,\textbf{S:}4.6469,\textbf{R:}64.0    &   \textbf{L:}1,\textbf{S:}2.3944,\textbf{R:}75.1    &   \textbf{L:}1,\textbf{S:}6.8179,\textbf{R:}60.8   \\
                           % & IP-greedy      &       &       &       &       &       \\
                           & PGS            &   \textbf{L:}11,\textbf{S:}4.8259,\textbf{R:}96.8    &   \textbf{L:}27,\textbf{S:}4.7658,\textbf{R:}85.4    &    \textbf{L:}47,\textbf{S:}4.6496,\textbf{R:}67.2   &     \textbf{L:}14,\textbf{S:}2.3944,\textbf{R:}74.8  &    \textbf{L:}42,\textbf{S:}6.8478,\textbf{R:}65.5   \\
                           & PDS            &   \textbf{L:}168,\textbf{S:}4.8014,\textbf{R:}89.7    &  \textbf{L:}213,\textbf{S:}4.7834,\textbf{R:}85.2     &   N/A    &   N/A    &    N/A   \\
                           & PSS            &   \textbf{L:}31,\textbf{S:}4.8261,\textbf{R:}98.1    &   \textbf{L:}60,\textbf{S:}4.7763,\textbf{R:}97.5    &   \textbf{L:}91,\textbf{S:}4.6736,\textbf{R:}96.2    &     \textbf{L:}29,\textbf{S:}2.4017,\textbf{R:}95.8  &    \textbf{L:}185,\textbf{S:}6.8869,\textbf{R:}96.4  \\
\bottomrule
\end{tabular}
\caption{Latency (L, ms), score (S), and recall rate (R, percentage) of all methods on three datasets.}
    \label{tab:main}
\end{table*}

We implemented all three proposed algorithms in C++: progressive greedy search (PGS), progressive degree search (PDS), and progressive score search (PSS). The code is provided in \url{https://github.com/DGENJI/Progressive-Search}.

\begin{figure}[!t]
  \centering
  \begin{minipage}[c]{0.51\textwidth}
		\centering
		\includegraphics[width=\textwidth]{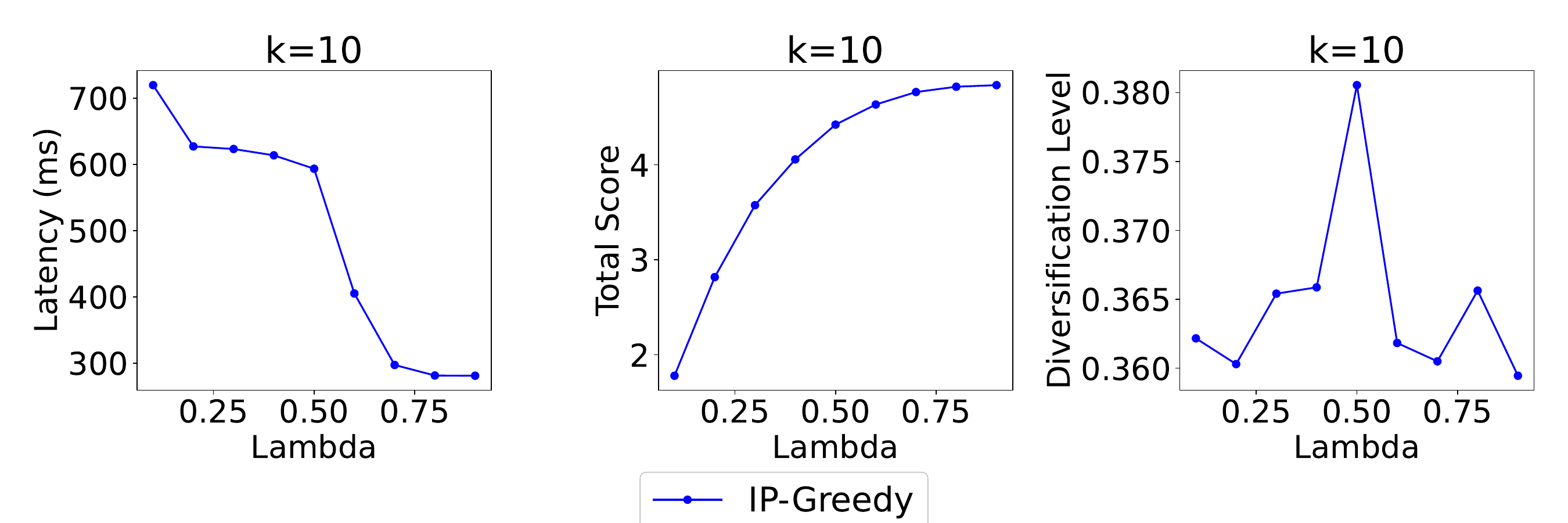}
        \caption{Results of IP-greedy on Txt2img when $k=10$}
        \label{fig:ip}
	\end{minipage} 
\end{figure}

Three baseline methods are implemented: div-A*, greedy, and IP-greedy. To better fit the AD$k$-NNS use case, we made several modifications to these baselines. For div-A*, the proximity graph is constructed during the online stage using the top $X$ nodes with the highest similarity scores, where $X$ is sufficiently large to ensure an optimal diverse set for all queries. However, this approach is impractical for real-world applications, as it assumes prior knowledge of all queries $Q = (q, k, \epsilon)$. For the greedy algorithm, we set the beam width to $L=400$, and assign a similarity score of 0 to any missing vectors if the result set contains fewer than $k$ elements. For IP-greedy, we use the code from \url{https://github.com/peitaw22/IP-Greedy}. Since it was originally designed for the $k$-MIPS problem, we only applied it to the LAION-art and Txt2img datasets, as their metric spaces are cosine similarity and inner product, respectively. However, during our experiments, we observed that for fixed values of $k$ and $c$, adjusting the parameter $\lambda$ does not significantly affect the diversification level. This finding is consistent with the experimental results reported in \cite{10.1145/3523227.3546779}. We present the experimental results when $k=10$ on Txt2img dataset separately in Figure~\ref{fig:ip}, and exclude IP-greedy from further comparisons with other methods.

All experiments were conducted on a Windows 11 system with 64GB of memory and an Intel Core i9-13900KF processor.

The motivation for this work typically arises from tasks that involve data in the cross-modalities. Therefore, we evaluated our methods on million-scale cross-modality datasets, including Deep1M, Txt2img, and LAION-art, which represent the three most common metric spaces: Euclidean, inner product, and cosine similarity. The consequently similarity function is defined as:
\begin{equation}
sim_{L2}(u,v) = 1-\sqrt{\sum_{i=1}^d (u_i-v_i)^2}
\end{equation}
\begin{equation}
sim_{ip}(u,v) = \sum_{i=1}^d u_i\times v_i
\end{equation}
\begin{equation}
sim_{cos}(u,v) = \sum_{i=1}^d \frac{u_i\times v_i}{\|u\|\times\|v\|}
\end{equation}
The dataset information is listed in Table~\ref{tab:dataset}.

We use HNSW as the proximity graph, adopting construction parameters: $M = 16$ and $ef = 200$ during the construction phase. In the search phase, we evaluate queries with $k = 5, 10, 15, 20$, and vary $\epsilon$ according to the characteristics of different datasets. The parameter of the efficiency level $ef$ is set from 10 to 100 in increments of 10, to demonstrate the trade-off between efficiency and the quality of the result.

Efficiency is measured by query latency, while result quality is evaluated by the total similarity score and recall rate, defined as follows:
\begin{equation}
\text{score} = \sum_{i=1}^k sim(r_i, q)
\end{equation}
\begin{equation}
\text{recall} = \frac{tp}{tp + fn}
\end{equation}
where $tp$ and $fn$ represent the numbers of true positives and false negatives, respectively. The ground truth is computed using the baseline div-A* algorithm.

\begin{figure}[!t]
  \centering
  \begin{minipage}[c]{0.51\textwidth}
		\centering
		\includegraphics[width=\textwidth]{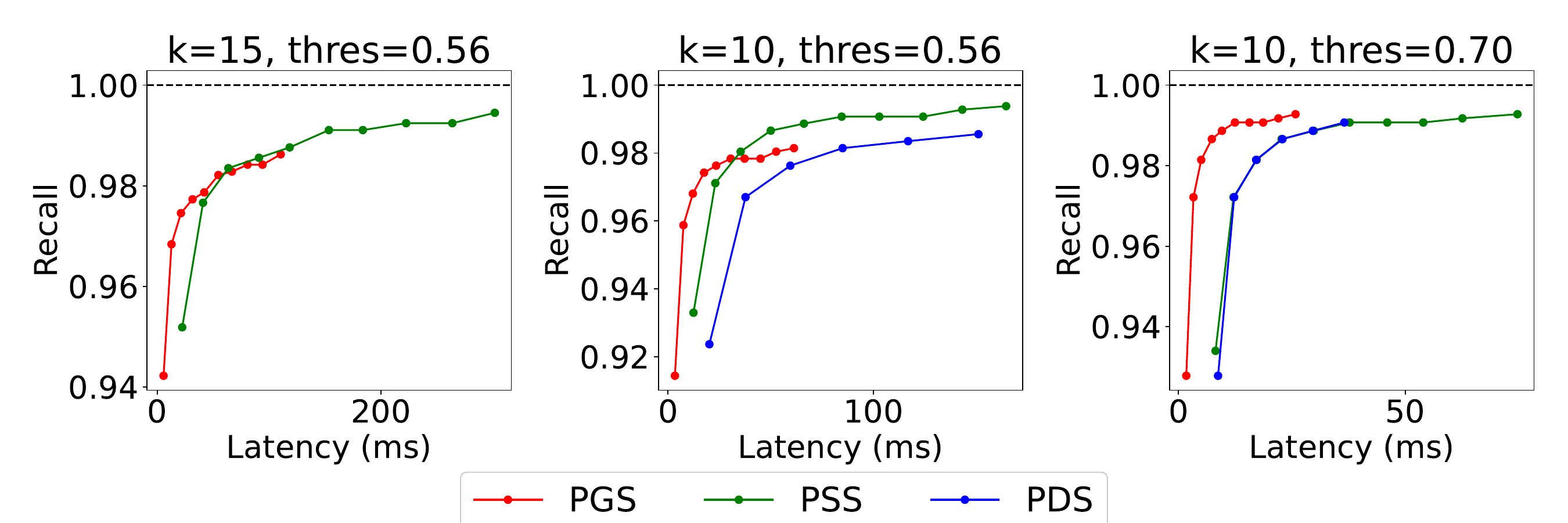}
        \caption{Latency-recall plots of Deep1M}
        \label{fig:deep}
	\end{minipage} 
    \begin{minipage}[c]{0.51\textwidth}
		\centering
		\includegraphics[width=\textwidth]{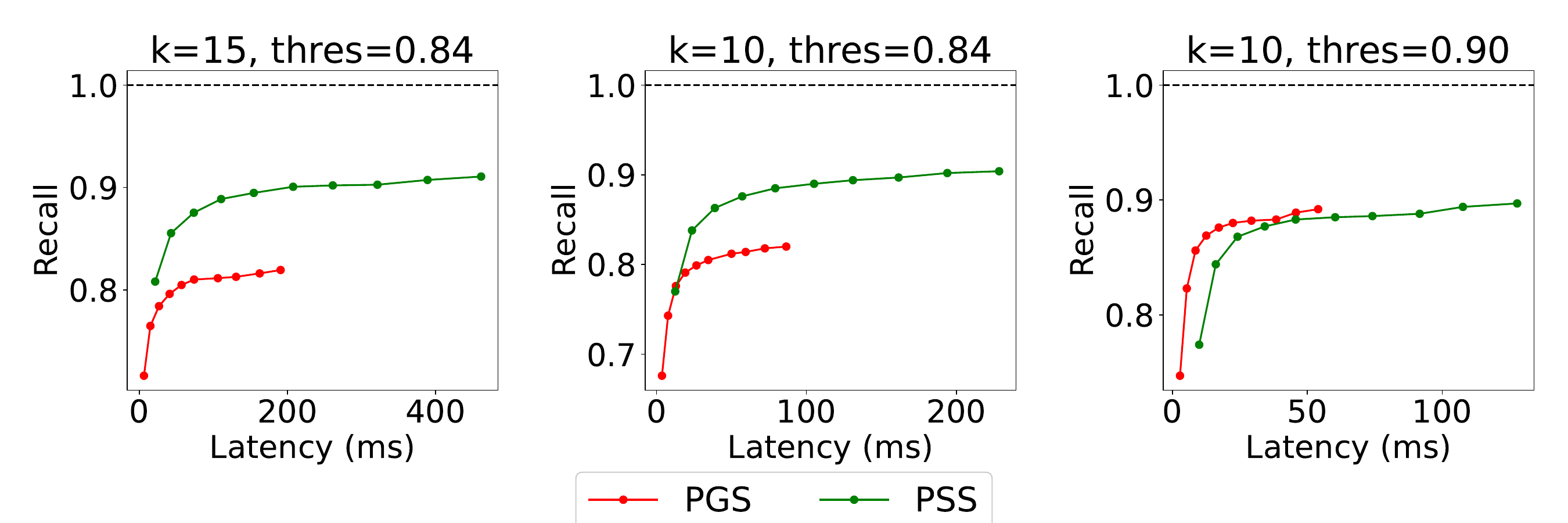}
        \caption{Latency-recall plots of LAION-art}
        \label{fig:laion}
	\end{minipage} 
    \begin{minipage}[c]{0.51\textwidth}
		\centering
		\includegraphics[width=\textwidth]{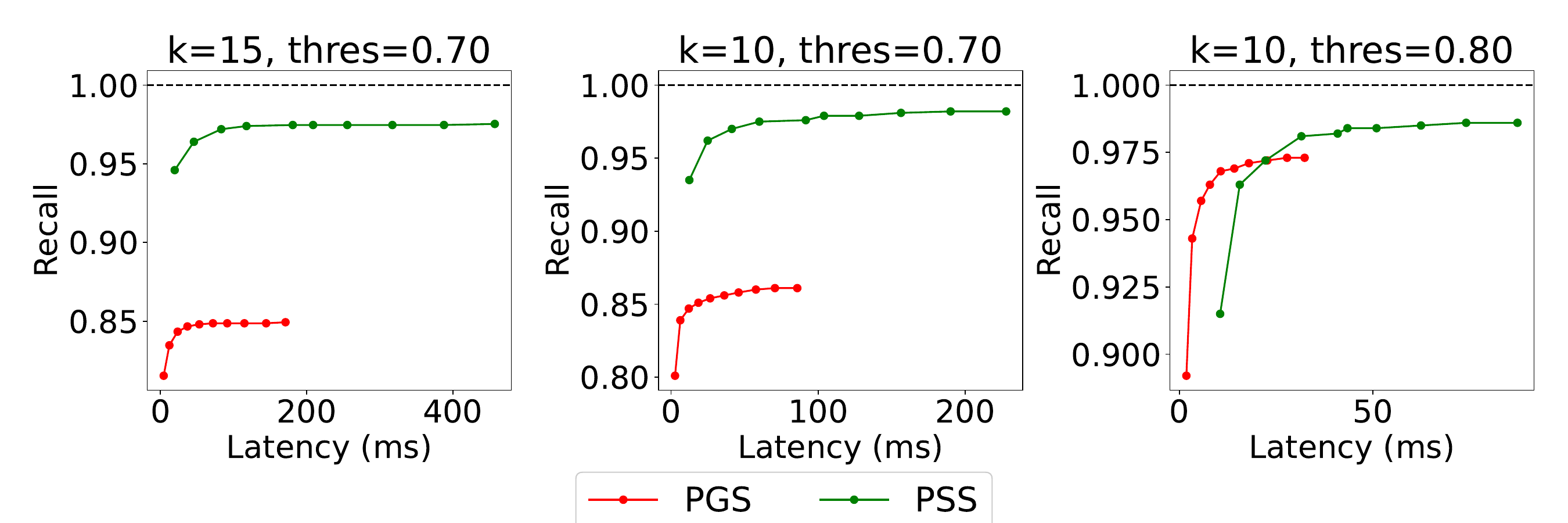}
        \caption{Latency-recall plots of Txt2img}
        \label{fig:txt2img}
	\end{minipage} 
\end{figure}

The primary experimental results are presented in Table~\ref{tab:main}, Figures~\ref{fig:deep_f}, \ref{fig:laion_f} and \ref{fig:txt2img_f}. In Table~\ref{tab:main}, we evaluated the methods under five experimental settings to provide a clear comparison of different diversification levels with a fixed $k$, and different $k$ values with a fixed diversification level. The high, medium, and low diversification levels are defined by $\phi(\epsilon) \approx 500, 100, 10$ in million-scale datasets (see Section~\ref{Analysis}). For the three progressive search methods, we report the results for a specific $ef$ value where both the recall rate and the score remain stable, defined as less than a $1\%$ change in the recall rate upon increasing $ef$. The "N/A" entries in the table indicate that the method's performance under those settings was too poor to provide a meaningful reference value, i.e., the latency is higher than the div-A* baseline. In Figures~\ref{fig:deep_f}, \ref{fig:laion_f}, and \ref{fig:txt2img_f}, we show how latency, score, and recall rate change as the similarity threshold $\epsilon$ varies, with $k=10$ and $ef=40$. The choice of $ef$ is based on the observation that progressive search methods converge around $ef=40$ (as shown later in Figures~\ref{fig:deep}, \ref{fig:txt2img}, and \ref{fig:laion}). Moreover, since $k \times ef = 400$, this makes the comparison with the greedy algorithm at $L=400$ more meaningful. Since PSS generally outperforms PDS, we only plot the curves for the greedy algorithm, PGS, and PSS in these figures. The latency of div-A* is too high to include in the threshold-latency plots, but we indicate the optimal quality achieved by div-A* with black lines.

To show a clear comparison among the proposed methods and a clear trade-off between latency and quality, we plot the latency-recall performance as the parameter $ef$ increases, under three settings for each dataset. The resulting plots are presented in Figures~\ref{fig:deep}, \ref{fig:txt2img} and \ref{fig:laion}. In addition to the three curves representing the PGS, PDS, and PSS algorithms, the optimal result computed by div-A* is indicated by a black dotted line. Since the sufficient estimation in PDS can be unnecessarily large, PDS performs poorly when $k$ and $\phi(\epsilon)$ are large. Therefore, we omit the PDS curve in cases where its latency is significantly higher than that of the other two methods.

The PDS and PSS algorithms can be regarded as approximation algorithms for the div-A* algorithm. To demonstrate the speedup achieved and to illustrate that PSS produces a tighter estimate of $K$ in vector databases compared to PDS, we present in Table~\ref{tab:k} the final value of $K$ obtained by each algorithm when $ef=10$ after the end of the query.

\subsection{Overall Evaluation}

\begin{figure}[!t]
  \centering
  \begin{minipage}[c]{0.51\textwidth}
		\centering
		\includegraphics[width=\textwidth]{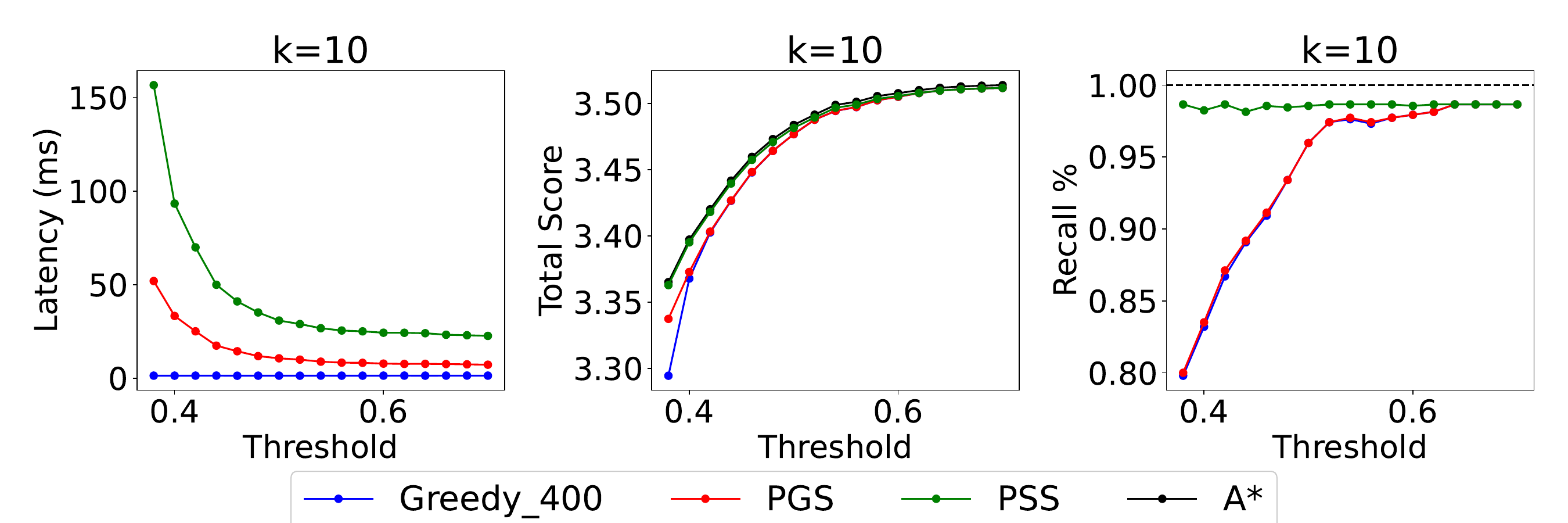}
        \caption{Results of methods on Deep1M when $k=10$}
        \label{fig:deep_f}
	\end{minipage} 
    \begin{minipage}[c]{0.51\textwidth}
		\centering
		\includegraphics[width=\textwidth]{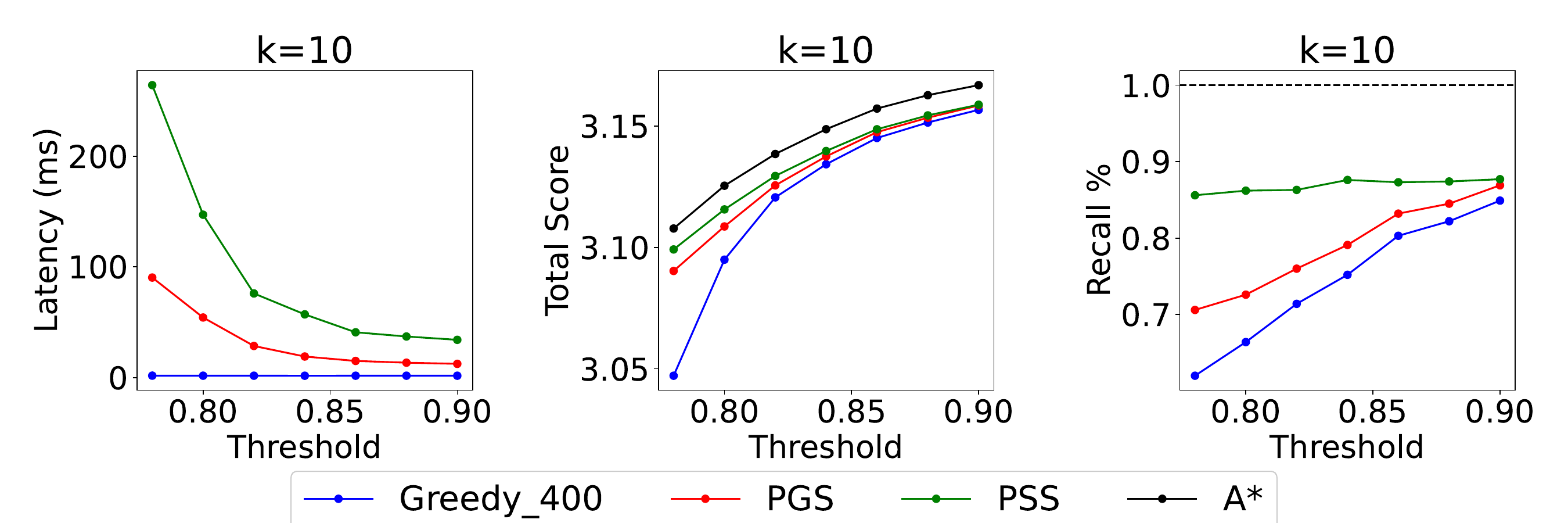}
        \caption{Results of methods on LAION-art when $k=10$}
        \label{fig:laion_f}
	\end{minipage} 
    \begin{minipage}[c]{0.51\textwidth}
		\centering
		\includegraphics[width=\textwidth]{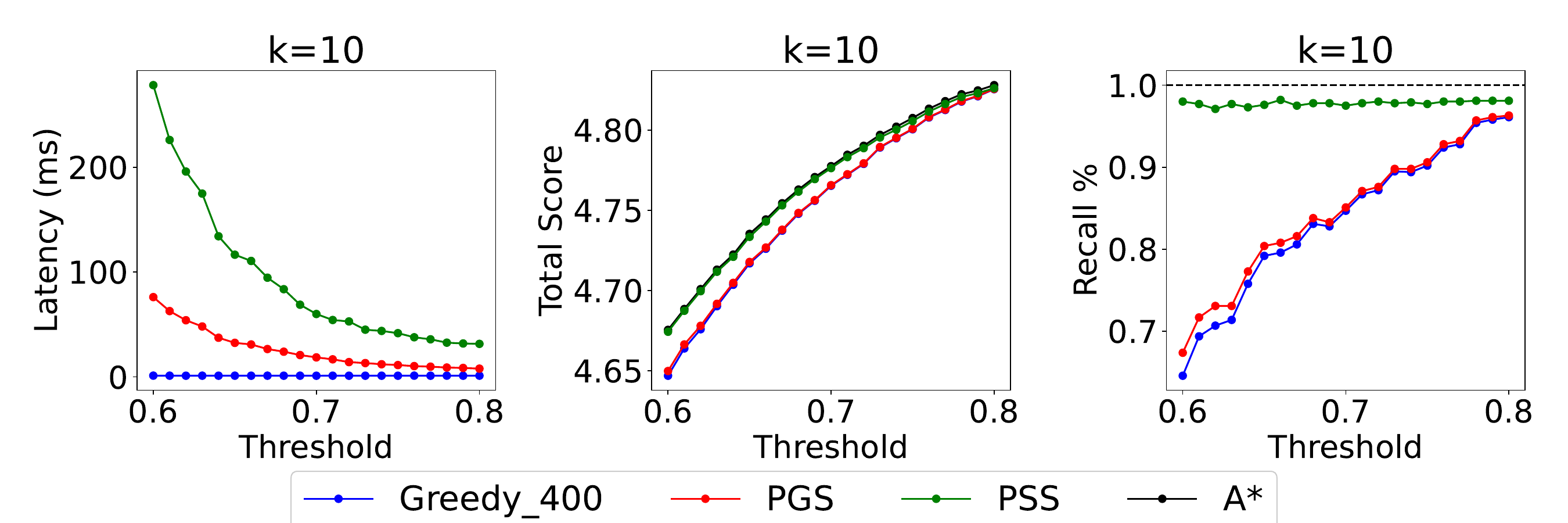}
        \caption{Results of methods on Txt2img when $k=10$}
        \label{fig:txt2img_f}
	\end{minipage} 
\end{figure}

Based on the observation of Table~\ref{tab:main}, Figures~\ref{fig:deep_f}, \ref{fig:laion_f} and \ref{fig:txt2img_f}, we draw the following conclusion with respect to the three evaluation criteria.

    \textbf{Latency}: Generally, the trend of latency is evident: higher values of $k$ and $\phi(\epsilon)$ result in increased latency. For A*-based algorithms, the proximity graph becomes denser and potentially larger in size. For progressive search methods, larger $K$ values require more time for beam search.

    The div-A* takes the longest time in most cases (except for the "N/A" entries). It is expected since it must compute $X$ similarity scores, build a proximity graph, and run the A* algorithm.
    
    The latency trend of greedy algorithm is exceptional: its latency is the lowest and does not vary much across different settings. This is because the workload of beam search and greedy result diversification are both fixed when $L$ and $k$ are both fixed. Greedy algorithm is thus extremely efficient, but it's limitation is shown later when we evaluate the result quality.

    The PDS exhibits strong performance for simpler queries. However, as $k$ and $\phi(\epsilon)$ increase, its latency becomes unacceptable, often surpassing that of the div-A* baseline. This is primarily due to an overestimated $K$ in these scenarios (see Table~\ref{tab:k}). We will discuss this in detail in Section~\ref{sec:al};

    Compared to the div-A* baseline, PGS and PSS achieve substantial speedups, reaching up to 42$\times$ at high diversification levels (see Deep1M, $k=15$, $\phi(\epsilon)\approx 500$). Although they are both considerably slower than the greedy approach, they can still process queries in milliseconds in most scenarios. PGS is generally 2$\times$ to 4$\times$ faster than PSS, because PGS is actually a component of PSS.
    
    % We also observe from the plots that for the same value of $ef$, PDS and PSS consistently require more time than PGS, because PSS performs additional computations compared to PGS, and PDS's diversification construction requires more computation than the greedy selection component of PGS. However, this increased latency is generally acceptable for three reasons: (1) The latency remains at the millisecond level, meaning it does not become a bottleneck in practical applications. (2) The latency of PSS with low $ef$ is still lower than PGS with high $ef$ in most cases.

    % However, PSS does have its limitations: its performance degrades when $k$ is large and $\epsilon$ is low (see $k=20$ and $\phi(\epsilon)\approx500$ cases). In such scenarios, the div-A* algorithm becomes a clear bottleneck, as its complexity is $O(K^k)$.
    
    \begin{table*}[]
    \centering
    \begin{tabular}{@{}lccccc@{}}
        \toprule
        \textbf{Algorithm} & $k=5, \phi(\epsilon)< 10$ & $k=20, \phi(\epsilon)<10$ & $k=5, \phi(\epsilon)\approx 100$ &$k=20, \phi(\epsilon)\approx 100$ \\ \midrule
        A* & L: 582ms, S: 1.8267 & L: 838ms, S: 6.7338& L: 834ms, S: 1.8214& L:851ms, S: 6.7066\\
        PDS & L: 23ms, S: 1.82603, R: 0.988 & L: 113ms, S: 6.73177, R: 0.991 & L: 36ms, S: 1.82123, R: 0.977 & L: 7010ms, S: 6.70635, R: 0.987\\
        PSS & L: 23ms, S: 1.82603, R: 0.988 & L: 114ms, S: 6.73182, R: 0.991 & L: 24ms, S: 1.82079, R: 0.988 & L: 166ms, S: 6.70475, R: 0.992\\ \bottomrule
    \end{tabular}
    \caption{Latency (L), score (S) and recall rate (R) of PDS, PSS and A* for Deep1M dataset}
    \label{tab:late}
\end{table*}

\begin{table*}[]
    \centering
    \begin{tabular}{@{}lcccccc@{}}
        \toprule
        \textbf{Algorithm} & $k=5, \phi(\epsilon)< 10$ & $k=20, \phi(\epsilon)<10 $ & $k=5, \phi(\epsilon)\approx 100$ &$k=20, \phi(\epsilon)\approx 100$& $k=5, \phi(\epsilon)\approx 500$ & $k=20, \phi(\epsilon)\approx 500$  \\ \midrule
        PDS & \textbf{AVG:}5.1, \textbf{MAX:}7 & \textbf{AVG:}20.1, \textbf{MAX:}24 & \textbf{AVG:}8, \textbf{MAX:}157 & \textbf{AVG:}341, \textbf{MAX:}10726 & \textbf{AVG:}3143, \textbf{MAX:}38383 & N/A\\
        PSS & \textbf{AVG:}5.3, \textbf{MAX:}10 & \textbf{AVG:}21, \textbf{MAX:}40 & \textbf{AVG:}6, \textbf{MAX:}30 & \textbf{AVG:}26, \textbf{MAX:}60 & \textbf{AVG:}18, \textbf{MAX:}230 & \textbf{AVG:}77, \textbf{MAX:}1020\\  \bottomrule
    \end{tabular}
    \caption{Comparison between the average (AVG) and maximum (MAX) estimated $K$ of PDS and PSS for Deep1M dataset when $ef=10$}
    \label{tab:k}
\end{table*}

    \textbf{Score}: It is clear that the total similarity score decreases as $\phi(\epsilon)$ increases, due to the stricter constraints imposed on the result set. When the diversification level is low ($\phi(\epsilon)<10$), all methods provide good approximations to the optimal score. However, when the diversification level is high ($\phi(\epsilon)\approx100,500$), there is a gap between the best performance of greedy-based methods and the optimal result, which cannot be closed simply by increasing $ef$. In contrast, our PSS method can closely approximate the optimal score when $ef$ is high for any $\epsilon$.
    
    Since the datasets consist of high-dimensional vectors, similarity scores are typically very close, leading to minimal proportional differences in the total scores between different methods (see Figures~\ref{fig:deep_f}, \ref{fig:laion_f} and \ref{fig:txt2img_f}). Even when the recall rates of two methods differ significantly (see $k=15$, $\phi(\epsilon)\approx 500$), the score difference remains below 3\%. Nevertheless, in most cases, higher recall rates correspond to higher scores. So, the score is only for reference. In the following analysis, we focus on the recall rate.

%     For all datasets, especially the LAION-art dataset, the optimal score is noticeably higher than the best score achieved by all methods. This is because HNSW does not return the accurate candidates for all queries. Take LAION-art as an example, since the embeddings all represent artwork, the dataset is partially dense. In such cases, a low score of $M=16$ in the proximity graph leads to poor connectivity. As a result, the accuracy of beam search remains low, regardless of how high $ef$ is set.
    
    \textbf{Recall}: When the diversification level is low ($\phi(\epsilon) < 10$), as discussed previously in Section~\ref{Analysis}, the optimal diverse set has substantial overlap with the top-$k$ results. In this scenario, the accuracy of the proximity graph is the most critical, and all methods achieve recall rates close to 1 when $ef$ is high. Notably, PSS still achieves the highest recall rate in these cases. Conversely, when the diversification level is high ($\phi(\epsilon)\approx100,500$), greedy and PGS approximate only the total similarity score rather than the optimal diverse set, resulting in a low recall rate that cannot be improved simply by increasing the accuracy of the proximity graph. In contrast, PSS employs the div-A* algorithm and achieves a recall rate close to 1 when $ef$ is high. For example for Txt2img $k=15$, $\phi(\epsilon)\approx500$ case, PSS has a recall rate of 96.4\%, while the PGS only has a recall rate of 65.5\%. It shows the capability of PSS to identify the optimal diverse set.

    Notice that for the LAION-art dataset, all methods struggle to achieve a recall rate of 1. This is because HNSW does not return the accurate candidates for all queries. In LAION-art dataset, since the embeddings all represent artwork, the dataset is partially dense in the vector space. In such cases, a low value of $M=16$ during the construction of proximity graphs leads to poor connectivity and slow convergence. As a result, the accuracy of the beam search remains low, regardless of how high $ef$ is set.

\subsection{Comparison Among Progressive Search Methods}
In this section, we focus on the analysis of plots in Figures~\ref{fig:deep}, \ref{fig:txt2img} and \ref{fig:laion}. For each row, the three plots are case $k=15$, $\phi(\epsilon) \approx100 $, $k=10$, $\phi(\epsilon) \approx100$ and $k=10$, $\phi(\epsilon) < 10$.

For the same value of $ef$, the result points for PSS consistently appear above and to the right of those for PGS in the latency-recall plots. Overall, the method shows a significant improvement when $ef=20$ compared to $ef=10$. However, the improvement becomes less pronounced as $ef$ increases further. In most cases, the recall rate reaches its maximum when $ef \approx 40$.

Changing $\epsilon$ and $k$, the graphs reveal the following trends:

\textbf{Easy Queries}: When the diversification level is low and $k$ is small, as discussed previously in Section~\ref{Analysis}, the optimal diverse set has substantial overlap with the top-$k$ results. In this scenario, accuracy is primarily determined by the efficiency of the beam search. Under these conditions, PGS generally outperforms the other methods, as its curve is typically above and to the left of the PDS and PSS curves. For example, to achieve a 99\% recall rate in the Deep1M dataset with $\epsilon=0.7$ ($\phi(\epsilon) < 10$) and $k=10$, PGS, PDS, and PSS require 10, 36, and 37 ms, respectively. Notably, the PDS and PSS curves coincide because $K \approx k$ in this case. While the improvement in result quality is not significant, the additional computation required by PDS and PSS compared to beam search is substantial.
    
 \textbf{Challenging Queries}: When the diversification level is high and $k$ is big, either the PGS and PSS curves overlap or the PSS curve appears above. In this scenario, the sub-optimality of the greedy-based algorithm becomes apparent. For example, on the Txt2img dataset with $\epsilon=0.7$ ($\phi(\epsilon) \approx100$) and $k=15$, PGS achieves maximum recall rates of 0.85, while PSS achieves recall rates of 0.97. In this case, PSS usually outperforms PDS by delivering higher-quality results within the same time constraints. On the Txt2img dataset with $\epsilon=0.7$ and $k=15$, when PGS achieves it's best recall of 0.85 in 37 ms, PSS achieves 0.95 in 19 ms and 0.97 in 45 ms.

\subsection{Comparison Among PDS, PSS and div-A*}\label{sec:al}

From an algorithmic perspective, both PDS and PSS can be regarded as approximation algorithms for the div-A* algorithm, as ANNS techniques inherently build an index that enables operations on a small subset of the dataset. To illustrate the speedup and quality trade-offs, supplementing Table~\ref{tab:main} for $k=20$, we summarize the latency, score, and recall rate across four experimental settings for the Deep1M dataset in Table~\ref{tab:late}.

As shown in Table~\ref{tab:late}, both methods consistently achieve recall rates close to 99\% across different settings, with speedups of up to 35$\times$ (PSS, $k=5$, $\phi(\epsilon)\approx 100$). However, PDS performs worse than div-A* when $k=20$ and $\phi(\epsilon)\approx 100$. This phenomenon can be attributed to two main reasons: (1) The estimation of $K$ is too large (see Table~\ref{tab:k}). Although $K < X$ for most easy queries, PDS computes for the most challenging queries a $K$ that is considerably larger than $X$. As a result, PDS spends excessive time on graph construction and the div-A* component for these queries. (2) The memory access pattern of beam search is random. So when $K$ is large, it may require much more time than a linear scan of the entire dataset.

\subsection{Estimation Of $K$}

As shown in Figures~\ref{fig:deep}, \ref{fig:txt2img}, \ref{fig:laion}, and Table~\ref{tab:main}, PSS achieves lower latency than PDS when the query is challenging. Examining Algorithms~\ref{alg:degree} and \ref{alg:score}, we observe that both methods require the construction of a diversity graph and the execution of the div-A* component. Notably, PSS includes an additional PGS component and may invoke the div-A* component multiple times. These results highlight the significant impact of a tight estimate of $K$ on overall performance. We report the estimations of $K$ for both methods under various settings in Table~\ref{tab:k}.

We can observe from Table~\ref{tab:k} that PDS only outperforms PSS when the diversification level is low ($\phi(\epsilon)< 10$). As $\epsilon$ decreases ($\phi(\epsilon)\approx 100,500$), the estimate provided by PDS becomes significantly larger, especially with large $k$. These phenomena support our analysis of the limitations of the two methods in Section~\ref{methodology}. Since our method is intended for scenarios that utilize vector databases with characteristics similar to those of our selected datasets, we can state with confidence that PSS is generally more effective.

Another important observation during our experiments is that, while most queries require $K$ to be only slightly larger than $k$, some challenging queries require much larger $K$, which can become a bottleneck. Identifying such difficult queries and optimizing them for them is an interesting and practical problem. However, since our methods are designed to address the AD$k$-NNS problem in general, this is beyond the scope of the present paper.
\section{Conclusion}
In this paper, we highlight the significance of the AD$k$-NNS problem, motivated by vector database applications such as recommendation system, RAG and image retrieval. To address this problem, we first propose progressive search, a specialized beam search framework designed to efficiently handle cases where the result size is initially uncertain. To further improve result quality, we incorporate the div-A* algorithm to compute the optimal solution of given candidates. Additionally, we introduce several theorems to establish an early stopping condition, enabling accurate estimation of the candidate set size and integrating seamlessly with the progressive search framework. Experimental results demonstrate that our method outperforms the baseline under medium to high diversification settings.
\section{AI-Generated Content Acknowledgement}
In the preparation of this paper, GPT-4.1 was used solely for polishing the language, enhancing clarity, and improving overall readability. All technical content, including methodology, experimental design, results, analyzes, and conclusions, was conceived by the authors without AI assistance.

Since the LAION-art dataset does not provide a query set of text embeddings, the authors used GPT-4.1 to randomly generate a portion of the query texts. These texts are later converted into embeddings by CLIP-ViT-
B-32.

\bibliographystyle{IEEEtran}
\bibliography{reference}

\end{document}